\documentclass[%
 twocolumn,
showpacs,
 amsmath,amssymb,fleqn
 aps,
prb,
floatfix,
]{revtex4}

\usepackage{graphicx}
\usepackage{bm}
\usepackage{color}
\usepackage{multirow}
\usepackage[mathlines]{lineno}

\newcommand{\vc}[1]{\ensuremath{\mathbf{#1}}}
\newcommand{\imp}{\ensuremath{{\rm imp}}}
\newcommand{\EF}{\ensuremath{E_{\rm F}}}
\newcommand{\mfpv}{\ensuremath{\boldsymbol{\Lambda}}}

\begin{document}


\title{Spin relaxation and spin Hall transport in 5{\it d} transition-metal ultrathin films}

\author{Nguyen H. Long} 
\email{h.nguyen@fz-juelich.de}
\affiliation{Peter Gr\"unberg Institut and Institute for Advanced Simulation, Forschungszentrum J\"ulich and JARA, D-52425 J\"ulich, Germany}
\author{Phivos Mavropoulos} 
\email{ph.mavropoulos@fz-juelich.de}
\affiliation{Peter Gr\"unberg Institut and Institute for Advanced Simulation, Forschungszentrum J\"ulich and JARA, D-52425 J\"ulich, Germany}
\author{Bernd Zimmermann, David S. G. Bauer, Stefan Bl\"ugel} 
\affiliation{Peter Gr\"unberg Institut and Institute for Advanced Simulation, Forschungszentrum J\"ulich and JARA, D-52425 J\"ulich, Germany}
\author{Yuriy Mokrousov} 
\affiliation{Peter Gr\"unberg Institut and Institute for Advanced Simulation, Forschungszentrum J\"ulich and JARA, D-52425 J\"ulich, Germany}

\begin{abstract}
  The spin relaxation induced by the Elliott-Yafet mechanism and the
  extrinsic spin Hall conductivity due to the skew-scattering are
  investigated in 5{\it d} transition-metal ultrathin films with
  self-adatom impurities as scatterers.  The values of the
  Elliott-Yafet parameter and of the spin-flip relaxation rate reveal a
  correlation with each other that is in agreement with the Elliott
  approximation.  At 10-layer thickness, the spin-flip relaxation time in
  5{\it d} transition-metal films is quantitatively reported about
  few hundred nanoseconds at atomic percent which is one and two
  orders of magnitude shorter than that in Au and Cu thin films,
  respectively.  The anisotropy effect of the Elliott-Yafet parameter
  and of the spin-flip relaxation rate with respect to the direction of 
  the spin-quantization axis in relation to the crystallographic axes is
  also analyzed.  We find that the anisotropy of the spin-flip relaxation
  rate is enhanced due to the Rashba surface states on the Fermi
  surface, reaching values as high as 97\% in 10-layer Hf(0001) film
  or 71\% in 10-layer W(110) film.  Finally, the spin Hall
  conductivity as well as the spin Hall angle due to the
  skew-scattering off self-adatom impurities are calculated using the
  Boltzmann approach.  Our calculations employ a relativistic version
  of the first-principles full-potential Korringa-Kohn-Rostoker Green
  function method.
\end{abstract}


\maketitle

\section{Introduction}

Spin-dependent transport phenomena in nanoscale structures such as
metallic thin films attract wide attention in spintronics where the
spin degree of freedom is manipulated for data transfer and storage in
information technology.\cite{wolf01,parkin02,zutic04} Due to spin-orbit 
coupling (SOC), an injected spin polarization in a metal decays exponentially 
in time as $\exp(-t/T_{\rm sf})$, where $T_{\rm sf}$ is the spin-flip relaxation time. 
Therefore, understanding and  manipulating the spin relaxation processes is one of the essential conditions for practical applications.
\cite{zutic04,johnson85,johnson88} To give two practical 
examples, in spin-information devices the spin-flip relaxation time
is usually required to be large.\cite{kimura05} On the contrary,
in ultrafast magnetization reversal devices, a short spin-relaxation
time is neccesary.\cite{beaurepaire96}
Spin-orbit induced scattering processes in metals are also at the origin of the spin Hall effect
(SHE),\cite{hirsch99,kato04,valenzuela06,seki08,gradhand10}
where a spin current is detected in the direction perpendicular to an
applied electric field,
or the inverse spin Hall effect (ISHE),\cite{saitoh06,kimura07} where
a spin current, injected into a nonmagnetic metal, induces a
transverse charge current. The SHE and ISHE have become effective ways for
spin current manipulation and detection in nano-devices.

It is well established that the Elliott-Yafet
mechanism\cite{elliott54,yafet62} of spin-flip scattering plays the
most important role in metals with time-reversal
symmetry\cite{kramers30} (i.e., non-magnetic) and
space-inversion symmetry.  Owing to the presence of spin-orbit
coupling, the Bloch wavefunctions are superpositions of the
spin-up and spin-down states which allow a spin-flip scattering off
impurities at low temperatures or off phonons at high temperatures
even if the scattering potential is spin-diagonal.
In the Elliott approximation, after neglecting the form of the
scattering potential, the spin-flip relaxation rate $T_{\rm sf}^{-1}$ is estimated to
be proportional to the spin-mixing, or Elliott-Yafet parameter (EYP), $b^2$.

The Elliott-Yafet spin-relaxation mechanism and the spin Hall current
induced by the scattering off impurities in bulk metals have been
already investigated within models as well as by first-principles
calculations. \cite{fabian98,zhukov08,fedorov08,gradhand09,gradhand11,heers11,heers12,zimmermann12}
However, little is known about these effects in metallic thin films with
thickness in the nanometer regime.  Owing to the breaking of
translational symmetry in thin films, many parameters have to be taken
into account, such as the thickness and the crystalline orientation of
the films.  Importantly, Rashba surface states\cite{rashba60,krupin05}
can be formed around the Fermi level and they were shown to enhance
the spin-flip relaxation rate.\cite{heers11,miyata11,long13} 
Thin film is a keyword for reducing the size of spintronics devices.
It also gives a flexibility in manipulating electron spins in the spin Hall experiments. 

In previous works\cite{long13,long13b} we have investigated in depth the spin-relaxation mechanism in noble-metal and W(001) ultrathin films. 
Our calculations revealed spin-relaxation mechanisms
that were brought about by the reduced dimensionality and that would
not be present in the bulk of these metals. 
For one thing we found\cite{long13b} that the free-electron-like Fermi surface of the
noble metals, when projected in the surface Brillouin zone of the
ultrathin film, cuts through the Brillouin zone edge producing
spin-flip hot spots that have only been reported in the case of
multivalent metals\cite{fabian98} so far. Additionally we analyzed the
surface states in the case of W(001) films \cite{long13} and found
that the Rashba character can strongly contribute to spin
relaxation; we also saw an oscillatory behavior of the spin relaxation as a function of the film thickness. In both cases\cite{long13b, long13} we also found
a considerable anisotropy of the EYP as well as spin-relaxation rate with respect to
the angle between the injected spin polarization and the film normal.

\begin{figure}
\includegraphics[scale=0.45,trim= 30 30 30 20,clip=true]{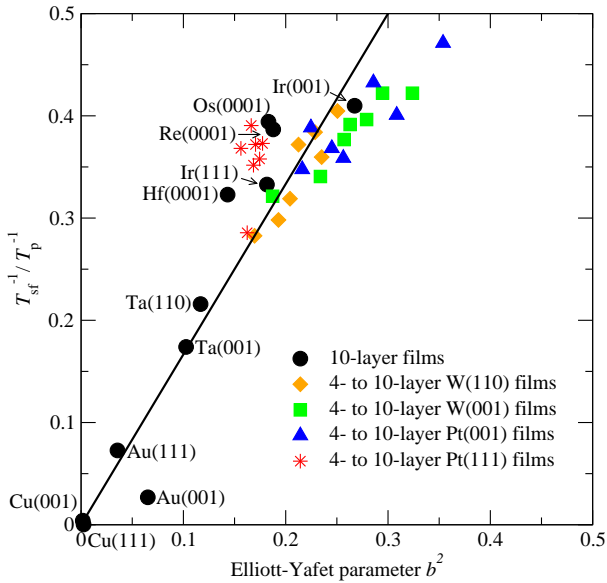} 
\caption{\small (Color online) Comparison between the ratio of the
  spin-flip rate and the momentum-relaxation rate for
  transition-metal films with self-adatom impurities and
  the Elliott-Yafet parameter. The solid black line is the function
  $f(x)=1.6x$ (fit by least squares) and serves as a guide to the
  eye.}
\label{comparebandspinrelax}
\end{figure}

In the present work we extend our computational study to a number of
metallic systems, namely ultrathin films of 5$d$ metals as well as Au
and Cu with different typical surface orientations: i.e. (111) and
(001) for fcc, (001) and (110) for bcc, and (0001) for hcp
structures. Our focus on 5$d$ metals is motivated by their strong SOC.
It is also well known that the 5{\it d} transition-metal surfaces such
as Pt, Ir, or W, are frequently used as substrates for growth of 3{\it
  d} magnetic thin films, while Cu, Pt or Au are frequently used as
conducting contacts or spin-Hall probes in spintronics experiments.
For the aforementioned systems we report on the calculation of the
Elliott-Yafet parameter, the spin-flip relaxation rate in the presence of
defect-induced electron scattering, and its anisotropy. Comparison
among different systems reveals that the spin-flip relaxation rate, and
in particular its anisotropy, has a spread of more than an order of
magnitude, even though the $5d$ systems are characterized by
comparable SOC strength and, to a crude approximation, by similarly
dense bands in the vicinity of the Fermi level. Additionally, we find that
the Elliott approximation to the spin-flip relaxation rate, that is
sometimes considered crude as it neglects the form of the scattering
potential, is qualitatively good in most cases.  We stress this
finding by demonstrating it already in Fig.~\ref{comparebandspinrelax}
where the ratio between spin-flip and momentum-relaxation rate,
$T_{\rm sf}^{-1}/T_{\rm p}^{-1}$, is plotted against the EYP for a
number of systems; these results are discussed in detail in
Sec.~\ref{sec:results}. Furthermore, we present calculations of the
spin Hall angle due to the extrinsic skew-scattering off defects.
This, too, can change in magnitude but also in sign depending on the
material, on the film thickness, and even on the surface orientation,
revealing the high complexity of the involved scattering processes.

\section{Theory}

The calculation of the electronic structure is done within the local
density approximation to density-functional theory in the
parametrization by Vosko \textit{et al}.\cite{vosko80} We employ the
full-potential Korringa-Kohn-Rostoker (FP-KKR) Green function
method\cite{Papanikolaou02,Stefanou90} as implemented in the
SPR-TB-KKR code\cite{SPR-TB-KKR} to calculate the self-consistent
electronic structure of the films and as implemented in the {\tt KKRimp}
impurity-embedding code\cite{bauer13} for the electronic structure of
the impurity adatoms.

Subsequently we calculate the Bloch wavefunctions, the Fermi surfaces,
and the scattering wavefunctions.  From the latter, the scattering
probability and the momentum- as well as the spin-flip relaxation rate can
be quantitatively determined via the spin-dependent scattering
matrix.\cite{fedorov08,gradhand09,gradhand11,heers11,heers12}
Knowledge of the spin-dependent scattering probability allows us to
employ the Boltzmann equation for spin Hall transport.\cite{gradhand10,mertig99,herschbach12}
Then, the spin Hall conductivity as well as the spin Hall angle
are calculated.  The formalism that we use is already given in
Ref.~\onlinecite{heers11,bauer13} and also partly in
Ref.~\onlinecite{zimmermann12,long13} of the authors.  The most
important expressions for making the present paper self-contained are
summarized in the following.

\subsection{The Elliott-Yafet parameter and the
  spin-flip scattering probability}

The metallic films that we treat here exhibit time-reversal invariance
(absence of external or internal magnetic fields) as well as
space-inversion invariance (that holds for finite-thickness films in
the bcc, fcc or hcp crystal structure).  Under these two
symmetry-invariant conditions there are two orthogonal degenerate
Bloch states at each {\bf k}-point in the band structure,
$\Psi^{+}_{\vc{k}}$ and $\Psi^{-}_{\vc{k}}$, which can be written as
superpositions of spin-up and spin-down states:
\begin{equation}
\begin{array}{lll}
\Psi^{+}_{\vc{k}}(\vc{r})=\left[\left.a_{\vc{k}}(\vc{r})\right|\left.\!\uparrow\right>+\left.b_{\vc{k}}(\vc{r})\right|\left.\!\downarrow\right>\right]e^{i\vc{kr}},\\
\mbox{} \\
\Psi^{-}_{\vc{k}}(\vc{r})=\left[\left.a^*_{\vc{-k}}(\vc{r})\right|\left.\!\downarrow\right>-\left.b^*_{\vc{-k}}(\vc{r})\right|\left.\!\uparrow\right>\right]e^{i\vc{kr}}.
\end{array}
\label{mixwavefunction}
\end{equation}
These two conjugate states show opposite spin polarization
$\vc{S}_{\vc{k}}^{\pm}:=\frac{\hbar}{2}\langle
\Psi^{\pm}_{\vc{k}}|\boldsymbol{\sigma}|\Psi^{\pm}_{\vc{k}}\rangle $,
i.e., $\vc S_{\vc{k}}^{+}= -\vc S_{\vc{k}}^{-}$. In an experiment a
spin-quantization axis (SQA) is defined by a unit vector $\hat{s}$
such that the injected spin population is polarized along
$\hat{s}$. This situation is formally described by taking linear
combinations of the two conjugate states at any $\vc{k}$ and forming
new $\Psi^{\pm}_{\vc{k}}$ such that $\hat{s}\cdot\vc{S}_{\vc{k}}^{+}$
is maximized. It is then assumed, within the Elliott-Yafet approach,
that the injected spins occupy these particular states
$\Psi^{+}_{\vc{k}}$, while the spin-relaxation process occurs due to
scattering from $\Psi^{+}_{\vc{k}}$ into $\Psi^{-}_{\vc{k}'}$. In this
basis where $\hat{s}\cdot\vc{S}_{\vc{k}}^{+}$ is maximized, the spin
polarization is related to the coefficients $a_{\vc{k}}(\vc{r})$ and
$b_{\vc{k}}(\vc{r})$ as follows:
\begin{eqnarray}
a_{\vc{k}}^2 := \int
\left|a_{\vc{k}}(\vc{r})\right|^2\,d^3r&=&\frac{1}{2}+\frac{1}{\hbar}|\vc{S}^{+}_{\vc{k}}|
,
\label{apara} \\
b_{\vc{k}}^2 :=\int \left|b_{\vc{k}}(\vc{r})\right|^2\,d^3r&=&\frac{1}{2}-\frac{1}{\hbar}|\vc{S}^{+}_{\vc{k}}|
\label{bpara}
\end{eqnarray}
where $|\vc{S}^{+}_{\vc{k}}|=\hat{s}\cdot\vc{S}^{+}_{\vc{k}}$ by the
construction of the particular basis $\Psi^{\pm}_{\vc{k}}$.
The Elliott-Yafet parameter $b_{\hat{s}}^2$ is defined as an average over the
Fermi surface (FS),
\begin{equation}
b^2_{\hat{s}}:=\left<b_{\vc{k}}^2\right>_{\rm
  FS}=\frac{1}{n\left(\EF\right)}\frac{1}{V_{\rm BZ}}\int_{\rm FS}\frac{d\vc{k}}{\hbar \left|\vc{v}_\vc{k}\right|}b_{\vc{k}}^2,
\label{bsq}
\end{equation}
where $\vc{v}_{\vc{k}}$ is the Fermi velocity and $n(\EF)$ is the
density of states at the Fermi level. The subscript $\hat{s}$
indicates that the value of $b_{\hat{s}}^2$ depends on the choice of
$\hat{s}$ through the dependence of $|\vc{S}^{+}_{\vc{k}}|$ (and thus
of $b_{\vc{k}}^2$). Thus we define the anisotropy of the EYP as
\begin{equation}
\mathcal{A}\left[b^2\right]=\frac{\mathrm{max}_{\hat{s}}\left(b^2_{\hat{s}}\right)-\mathrm{min}_{\hat{s}}\left(b^2_{\hat{s}}\right)}{\mathrm{min}_{\hat{s}}\left(b^2_{\hat{s}}\right)}.
\label{aniso1}
\end{equation}
As we have found in previous works,\cite{zimmermann12,long13,long13b,mokrousov13}
depending on the material, $\mathcal{A}\left[b^2\right]$ can reach
large values, well exceeding 100\%.

Within the Elliott approximation, where the form of the scattering
potential is neglected and $b^2$ is assumed to be small, the spin-flip
probability $P^{+-}_{\vc{k}\vc{k'}}$ is approximately proportional to 
$b_{\vc{k}}^2$.  As a result, the ratio between the spin-flip relaxation rate $T_{\rm sf}^{-1}$ and the momentum-relaxation rate $T_{\rm p}^{-1}$ is
proportional to the EYP, $T_{\rm sf}^{-1}/T_{\rm p}^{-1}\propto b^2$.
This value depends on the electronic structure and the strength of
spin-orbit coupling of the materials.  Therefore, as we discuss later,
it varies from Hf to Pt in 5{\it d} group. For $5d$ transition metals
with adatom defects the assumptions of the Elliott approximation are
certainly not valid. Still, the proportionality $T_{\rm
  sf}^{-1}/T_{\rm p}^{-1}\propto b^2$ holds qualitatively, as we
discuss in Sec.~\ref{sec:results} and show in Fig.~\ref{comparebandspinrelax}.

\subsection{Scattering off impurities}

Now we employ the scattering matrix to calculate the spin relaxation
due to the impurity scattering.  We use indices $\sigma,\sigma'\in
\{+,-\}$ corresponding to the Bloch wavefunctions
$\Psi^{\pm}_{\vc{k}}$ of Eq.~(\ref{mixwavefunction}). The
wavefunctions scattered by the impurity at energy $E=E(\vc{k})$,
$\Psi^{\imp,\sigma}_{\vc{k}}(\vc{r})$, are calculated in terms of the
unscattered Bloch wavefunctions of the host via the Lippmann-Schwinger
equation
\begin{eqnarray}
  \Psi^{\imp,\sigma}_{\vc{k}}(\vc{r})&=&\Psi^{\sigma}_{\vc{k}}(\vc{r})\nonumber\\
&+&\int d^3r'G(\vc{r},\vc{r'};E)\Delta V(\vc{r'})\Psi^{\imp,\sigma}_{\vc{k}}(\vc{r'}).
\label{impwavefunc}
\end{eqnarray}
The wavefunctions appearing here are column-vectors in spin space.
The host Green function $G(\vc{r},\vc{r'};E)$ is a $2\times2$ matrix
in spin space. The same holds for $\Delta{V}$, the difference between
the impurity potential $V^{\imp}$ and the host potential including the
difference of the spin-orbit contributions.  The scattering matrix 
can be simply written in terms of the host and scattered wavefunctions
\begin{equation}
T^{\sigma\sigma'}_{\vc{k}\vc{k'}}=
\int d^3r\,[\Psi_{\vc{k}}^{\sigma}(\vc{r})]^{\dag}\,
\Delta V(\vc{r})\,
\Psi_{\vc{k}'}^{\imp,\sigma'}\!(\vc{r}).
\label{Tmatrix}
\end{equation}
The integration in Eqs.~(\ref{impwavefunc}) and (\ref{Tmatrix}) is
numerically confined in the atomic cells where the difference in
potential is found to be non-negligible.  Under assumption of elastic
scattering, the scattering probability due to a number of impurities
in the system is determined by the Golden Rule
\begin{equation}
P_{\vc{kk'}}^{\sigma\sigma'}=\frac{2\pi}{\hbar}Nc\left|T_{\vc{kk'}}^{\sigma\sigma'}\right|^2\delta\left(E_{\vc{k}}-E_{\vc{k'}}\right),
\label{probability}
\end{equation}
where $N$ is the number of atoms in the system and $c$ is impurity
concentration.  The linear dependence of
$P_{\vc{kk'}}^{\sigma\sigma'}$ on the number of impurities $cN$
implies that the scattering events are independent to each other and it is
expected to hold in the dilute concentration limit where the defects
do not form impurity bands.  The {\bf k}-dependent
relaxation rate can be calculated by summation over all $\vc{k'}$:
\begin{equation}
\left(\tau^{\sigma\sigma'}_{\vc{k}}\right)^{-1}=\sum_{\vc{k'}}P_{\vc{kk'}}^{\sigma\sigma'}=\frac{2\pi
  Nc}{V_{\rm BZ}}\int_{\rm FS}\frac{d\vc{k}'}{\hbar^2 \left|\vc{v}_{\vc{k'}}\right|}\left|T^{\sigma\sigma'}_{\vc{k}\vc{k'}}(\EF)\right|^2.
\label{krelaxtime}
\end{equation}
The relaxation rate averaged over the Fermi surface is obtained as
\begin{equation}
\left(\tau^{\sigma\sigma'}\right)^{-1}=\frac{1}{n(\EF)}\frac{1}{V_{\rm BZ}}\int_{\rm
  FS}\frac{d\vc{k}}{\hbar \left|\vc{v}_{\vc{k}}\right|}\left(\tau^{\sigma\sigma'}_{\vc{k}}\right)^{-1}.
\label{relaxtime}
\end{equation}
In a non-magnetic system, it is obvious that
$\left(\tau^{++}\right)^{-1}=\left(\tau^{--}\right)^{-1}$ which is the
spin-conserving relaxation rate $T_{\rm c}^{-1}$ and
$\left(\tau^{+-}\right)^{-1}=\left(\tau^{-+}\right)^{-1}$ which is the
spin-flip relaxation rate $T_{\rm sf}^{-1}$. 
The momentum-relaxation rate $T_{\rm p}^{-1}$ is then defined as
$T_{\rm p}^{-1}=T_{\rm c}^{-1}+T_{\rm sf}^{-1}$ and the spin-relaxation rate
$T_1^{-1}$ is defined as two times the spin-flip relaxation rate
$T_1^{-1}=2T_{\rm sf}^{-1}$.  The factor 2 appears since $T_1$ is
experimentally derived from the full linewidth at half-amplitude of
conduction electron resonance spectra.  

Similar to the anisotropy of the EYP, Eq.~(\ref{aniso1}), we have a definition of the anisotropy of
the spin-flip relaxation rate
\begin{equation}
\mathcal{A}\left[T_{\rm sf}^{-1}\right]=\frac{\mathrm{max}_{\hat{s}}T_{\rm sf}^{-1}(\hat{s})-\mathrm{min}_{\hat{s}}T_{\rm sf}^{-1}(\hat{s})}{\mathrm{min}_{\hat{s}}T_{\rm sf}^{-1}(\hat{s})}.
\label{aniso2}
\end{equation}

To attest the numerical accuracy of the calculation of the relaxation time, the optical theorem:
\begin{equation}
-\frac{2Nc}{\hbar}\mathrm{Im}T_{\vc{k}\vc{k}}^{\sigma\sigma}=\frac{2\pi
  N^2c}{V_{\rm BZ}\hbar}\sum_{\sigma'}\int_{\rm FS}\frac{d\vc{k}'}{\hbar \left|\vc{v}_{\vc{k'}}\right|}\left|T_{\vc{kk'}}^{\sigma\sigma'}(\EF)\right|^2
\label{opticaltheorem}
\end{equation}
is also checked. 
In our calculations of thin metallic films, the optical theorem is
very sensitive and is satisfied in most cases to within 5\% and in few infavorable cases, such as 4-layer Pt(111) or 4-layer Os(0001), to within 10\%. 

\subsection{Spin Hall conductivity}

To deal with the extrinsic spin Hall conductivity due to the skew-scattering off impurities, the Boltzmann equation is utilized.  
The method was successfully applied to investigate the spin Hall effect in Cu and Au bulk as well as in Au(111) thin films with various impurities.\cite{mertig99,gradhand10,herschbach12} 
Following Refs.~\onlinecite{mertig99,gradhand10}, we start from the
lineared Boltzmann equation for the mean free path $\mfpv$
\begin{equation}
\mfpv^{\sigma}(\vc{k})=\tau^{\sigma}_{\vc{k}}\left[{\vc{v}}_{\vc{k}}+\sum_{\vc{k'}\sigma'}P_{\vc{k'k}}^{\sigma'\sigma}\mfpv^{\sigma'}(\vc{k'})\right],
\label{lambda}
\end{equation}
where $P_{\vc{kk'}}^{\sigma\sigma'}$ is the scattering probability defined in Eq.~(\ref{probability}) and the relaxation time $\tau_{\vc{k}}^{\sigma}=1/\sum_{\sigma'}\left(\tau_{\vc{k}}^{\sigma\sigma'}\right)^{-1}$ which is calculated from Eq.~(\ref{krelaxtime}).

The term $\sum_{\vc{k'}\sigma'}P_{\vc{k'k}}^{\sigma'\sigma}\mfpv^{\sigma'}(\vc{k'})$ is called the scattering-in term and it can be separated into two parts: spin-conserving part when $\sigma'=\sigma$ and spin-flip part when $\sigma'\neq\sigma$.
For the 5{\it d} materials that have the strong SOC, the spin-flip part cannot be neglected. 

After self-consistently solving Eq.~(\ref{lambda}), the charge
conductivity tensor $\underline{\kappa}$ as well as the spin
conductivity tensor $\underline{\kappa}^{\rm s}$ are dertermined as 
\begin{equation}
\underline{\kappa}=\frac{e^2}{\hbar}\frac{1}{(2\pi)^2d}\sum_{\sigma}\int_{\rm
  FS}\frac{d\vc{k}}{\left|\vc{v}_{\vc{k}}\right|}\vc{v}_{\vc{k}}  \otimes\mfpv^{\sigma}(\vc{k})
\label{eq:ccond}
\end{equation}
and
\begin{equation}
\underline{\kappa}^{\rm s}=\frac{e^2}{\hbar}\frac{1}{(2\pi)^2d}\sum_{\sigma}\int_{\rm FS}\frac{d\vc{k}}{\left|\vc{v}_{\vc{k}}\right|}\left(\frac{2}{\hbar}S^z_{\vc{k}}\right)\ \vc{v}_{\vc{k}} \otimes\mfpv^{\sigma}(\vc{k}),
\label{eq:scond}
\end{equation}
respectively.  
$S^z_{\vc{k}}$ is the spin-expectation value calculated in Eq.~\ref{apara} and ~\ref{bpara} by choosing the spin-polarization direction along the film normal meaning that the spin-quantization axis $\hat{s}$ is taken along $z$-direction.
The expression for conductivity~(\ref{eq:ccond}-\ref{eq:scond}) takes into account the film thickness $d$ and can be directly compared to the conductivity in bulk. 
It is obvious that the charge and spin conductivity are inversely proportional to the impurity concentration.
However, the ratio between them, the spin Hall angle
$\alpha=\kappa^s/\kappa$, is independent of the impurity
concentration.

\subsection{General computational details}

In our calculations, an angular momentum cut off of $l_{\mathrm
  {max}}=3$ is taken and the experimental lattice parameters are used
for all elements.  In order to obtain impurity wavefunctions,
Eq.~(\ref{impwavefunc}), within the FP-KKR Green function method, the
impurity potential is calculated using the J\"ulich KKR
impurity-embedding code ({\tt KKRimp}). \cite{bauer13} The charge- and
spin-density screening of the impurity are self-consistently calculated
within a cluster of nearest neighbors of the impurity atom.  To test
the influence of the cluster size, larger clusters of up to 4th nearest
neighbors are also considered in some cases showing negligible
differences.  
    
\subsection{System of coordinates}

For definiteness we state here that throughout the paper we use the Cartesian structure coordinates of thin films $xyz$ with respect to the bcc, fcc or hcp structure basis. 
The following convention is used: $z$-axis is always the film normal and $x$- and $y$-axis are defined as the Cartesian coordinate related to $z$-axis, i.e. in fcc (001), bcc (001) or hcp (0001) films $x$- and $y$-direction are [100] and [010], respectively; in bcc (110) films, $x$- and $y$-direction are [001] and [1$\bar 1$0], respectively; in fcc (111) films, $x$- and $y$-direction are [1$\bar 1$0] and [$\bar 1$0$\bar 1$], respectively.

\section{Results and discussion\label{sec:results}}

In order to systematically investigate the spin relaxation of 5{\it d}
transition-metal thin films, we first calculate the EYP and discuss
the results in subsection \ref{secEYP}.  In subsection
\ref{secspinrelax}, the momentum-relaxation time and spin-flip relaxation time due to
scattering off self-adatom impurities are quantitatively analyzed.
In subsection \ref{secspinhall}, the spin Hall conductivity and spin Hall angle are studied.  
In order to compare results with free-electron-like metals, we also consider 
 Au(111), Au(001), Cu(111) and Cu(001) thin films. 
We examine the behavior of the calculated quantities with respect to film
thickness and orientation, and analyze them with respect to the Fermi
surface, in particular concerning the surface states.
 
\subsection{Elliott-Yafet parameter \label{secEYP}}

\begin{figure}
\includegraphics[scale=0.34,trim= 40 1 20 20,clip=true,angle=270]{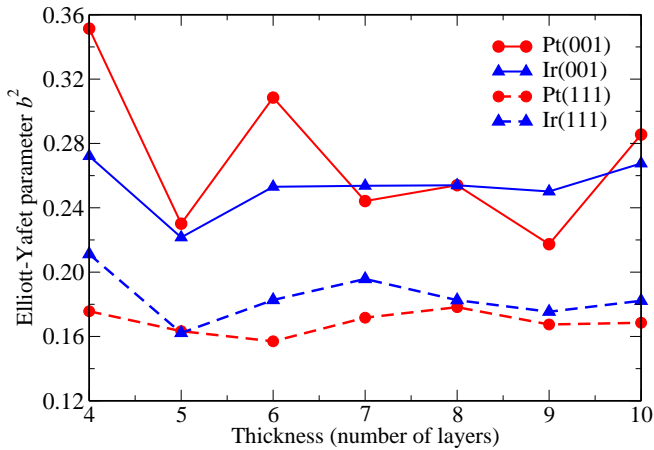}
\caption{\small (Color online) The thickness dependence of the Elliott-Yafet parameter in Pt(001), Pt(111), Ir(001) and Ir(111) thin films with the spin-quantization axis perpendicular to the films.}
\label{bsqPtIr}
\end{figure}
According to the Elliott approximation, the calculation of the EYP can
preliminarily describe the spin relaxation of the host materials.  In
a recent paper\cite{long13} we analyzed the EYP for bcc W(001) and
discussed in detail its dependence on the film thickness.  Remarkably,
we found that owing to the surface states at the Fermi surface, the
EYP of W(001) exhibits an oscillatory behavior with respect to the
film thickness, which we traced back to the interaction of surface
states at the two surfaces of the film together with the stacking of
the bcc structure.  In addition, the anisotropy of the EYP for W(001)
thin films was found to have a high value of 37\% at 10 layers film
thickness.  In the present work, we present analogous calculations of
the EYP for other 5{\it d} transition-metal thin films from Hf to Pt
with different crystal structures.

First we examine the film-thickness dependence of the EYP.  In
Fig.~\ref{bsqPtIr}, we show the calculations of the EYP as a
function of the film-thickness for fcc Pt(001), Pt(111), Ir(001) and
Ir(111) thin films with the SQA perpendicular to the films.  These
films are chosen since their EYPs show a variation with increasing the
film thickness.  In these systems, however, there are no surface
states at the Fermi surface, which cause a pronounced oscillatory
behavior of the EYP as found in W(001) films.\cite{long13} Yet the
influence from the stacking symmetry along $z$-direction,
i.e. ...ABAB... in fcc (001) films and ...ABCABC... in fcc (111) fims,
could give rise the fluctuation of the EYP as seen in Pt(001) and Pt(111).  
In the other 5{\it d} films, such as Ta(001), W(110) or the hcp-metal (0001) surfaces, the variation of the EYP as a function of the film thickness is much smaller.
 
Shifting our attention to the EYP at a certain thickness of Pt and Ir
films with the same crystalline orientation, we find that they are
quantitatively of the same order.  For instance, for 10-layer films,
the EYP of Pt(001) and Ir(001) has a value of 0.286 and 0.268,
respectively.  In 10-layer Pt and Ir (111)-films, they are both
smaller than those of (001) thin films but have similar values of
0.168 and 0.182, respectively.  To clearly see this trend, in the
first column of Table~\ref{alldata}, the values of the EYP for 5{\it d} films in
10-layer thickness  are summarized.  
It is also seen that the EYP of 0.184 for 10-layer hcp Os(0001) is very close to that of 0.187 for Re(0001). 

\begin{figure}
\includegraphics[scale=0.37]{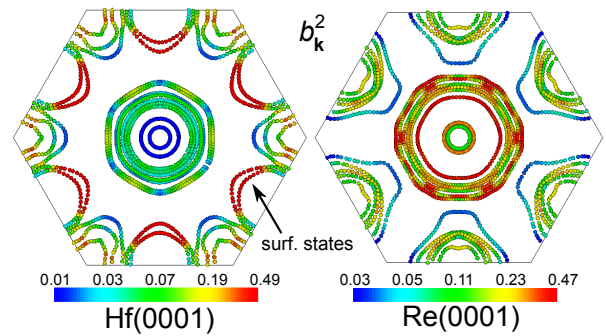} 
\caption{\small (Color online) The distribution of $b_{\vc{k}}^2$ on
  the Fermi surfaces of 10-layer films of Hf(0001) (left) and Re(0001)
  (right). One of the surface states of Hf(0001) is pointed at by an
  arrow. The SQA is perpendicular to the films.}
\label{bsqkHfRe}
\end{figure}  

For systems in which the Rashba surface states play a role, the EYP is very large. 
For example, among the bcc thin films, W(001) and W(110) that have surface states at \EF\ have much higher EYP as compared to Ta(001) and Ta(110) which do not
have the surface states.  
In 10-layer Hf(0001) the surface states at the Fermi surface also manifest in a large $b^2$ of 0.143 which is of the same magnitude as that of Os(0001) and Re(0001) films.  
To clarify the effect of the Rashba surface states, in Fig.~\ref{bsqkHfRe}, the distribution of $b_{\vc{k}}^2$ on
the Fermi surface of 10-layer Hf and Re (0001) films is examined.
One of the surface states of Hf(0001) is denoted by an arrow.  It is
obvious that in 10-layer Hf(0001) film the regions which provide a large contribution to 
$b^2$ are
mainly distributed over the surface states.  On the other hand, in 10-layer Re(0001) film, the
complicated electronic structure with many crossing bands causes many
spin-flip hot spots in the bulk-like states.  The surface
states-dependent effect is explained in
Refs.~\onlinecite{long13,miyata11} and one can apply the same
arguments to state that the EYP is enhanced due to the existence of
the Rashba surface states.

The EYP of 10-layer Au(111) and Au(001) films is one order of
magnitude smaller, and the EYP of 10-layer Cu(111) and Cu(001) is even
two orders of magnitude smaller than that of 5{\it d}
transition-metals.  This demonstrates the important role of {\it
  d}-states for spin-flip scattering promoted by strong mixing between
spin-up and spin-down states. The complicated electronic structure
with many band crossings in 5{\it d} thin films results in a large
density of spin-flip hot spots which considerably enhance the
spin-mixing parameter.  One can also observe that the surface states of
Au(111) and Cu(111) films have only a small contribution to the DOS
and thus contribute only little to the EYP.

\begin{figure}
\includegraphics[scale=0.37,trim= 1 1 1 1,clip=true]{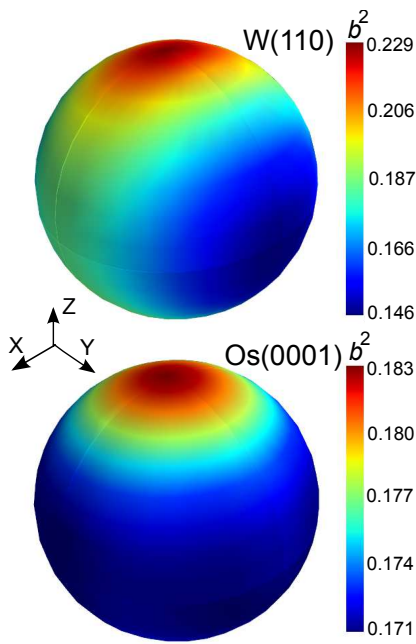} \\
\caption{\small (Color online) The value of $b^2$ of 10-layer W(110)
  and Os(0001) films for the spin-quantization axis $\hat s$ on the
  unit sphere.}
\label{anibsqWOs}
\end{figure}  
Now we investigate the effect of the anisotropy of the EYP with
respect to the SQA $\hat s$.  In Fig.~\ref{anibsqWOs}, the EYP is
plotted for $\hat s$ on the unit sphere for 10-layer W(110) and
10-layer Os(0001) films, chosen as two opposite extremes among our data, in which the former shows a
large value of anisotropy and the latter shows a small value.
The EYP of a 10-layer W(110) film varies in a large range
from 0.146 to 0.229,  while the EYP of a 10-layer Os(0001) film
varies in a smaller range from 0.171 to 0.183.  As a result, an
anisotropy $\mathcal{A}\left[b^2\right]$ of 57\% for W(110) and 7\%
for Os(0001) is found.  In addition, we calculated the anisotropy
value for different film-thickness.  The results show for all
systems that it is quite robust with respect to the film thickness.

\begin{figure}
\includegraphics[scale=0.37]{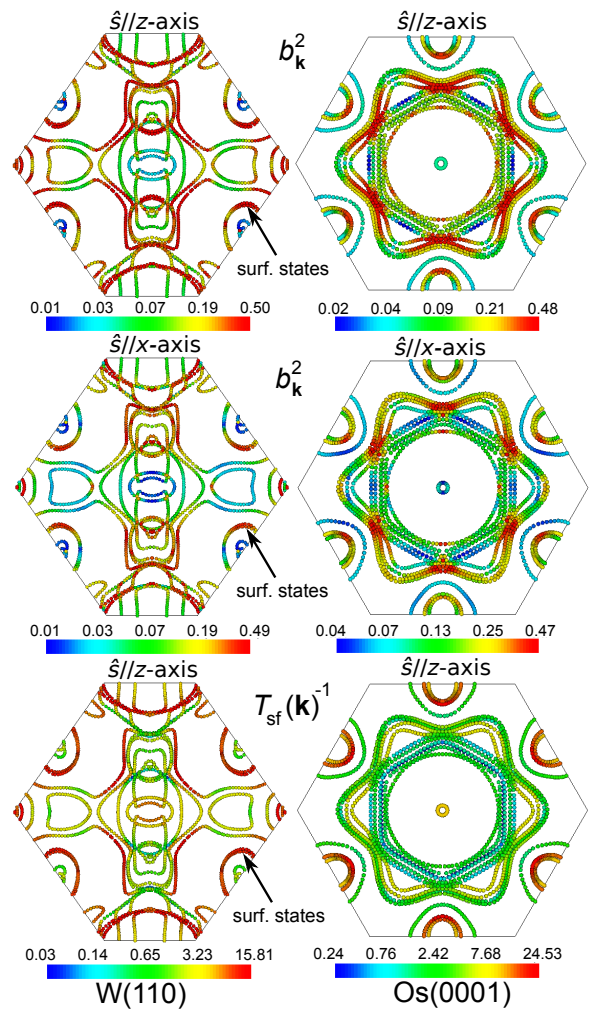} 
\caption{\small (Color online) At the top and in the middle: the distribution
  of $b_{\vc{k}}^2$ on the Fermi surfaces of 10-layer W(110) (left) and
  Os(0001) films (right) for the SQA along $z$-axis (top) and $x$-axis (middle). At the bottom: the distribution of $T_{\rm sf}(\vc{k})^{-1}$ on
  the Fermi surface of 10-layer W(110) film with W adatom defect
  (left) and 10-layer Os(0001) film with Os adatom defect (right) for
  the SQA along $z$-axis.}
\label{bsqandT_WOs}
\end{figure}  
Only part of the difference between the two metals is due to surface
states. In Fig.~\ref{bsqandT_WOs} the distribution of $b_{\vc{k}}^2$
on the Fermi surface of 10-layer W(110) film and 10-layer Os(0001)
film is shown for the SQA along the $z$- and $x$-direction.  One
can see that, by rotating the SQA from $z$-axis to $x$-axis, the
distribution of EYP in 10-layer W(110) film changes considerably not
only at the surface states but also at the bulk-like states.  On the
contrary, the hot spots at the Fermi surface of 10-layer Os(0001) film
remain when rotating the SQA.

In Refs.~\onlinecite{zimmermann12,long13}, we pointed out that the
reduction of symmetry in thin films, compared to the bulk of cubic
systems, will play a role for the anisotropy of the spin relaxation.
One clear evidence for this can be seen in Fig.~\ref{anibsqWOs} in which
the EYP of W(110) is maximal when $\hat s$ is parallel to the film normal.  
Moreover, the symmetry of EYP exactly corresponds to the crystallographic
symmetry.\cite{zimmermann12} 
The anisotropy of the EYP of 57\% in 10-layer W(110) that we
find here is much larger as compared to 6\% in bulk W.\cite{zimmermann12}
The anisotropy values
$\mathcal{A}\left[b^2\right]$ for other 10-layer films are summarized
in the second column of Table ~\ref{alldata}.  Similar to W(110),
other cubic films show a relatively high anisotropy value of the EYP
as compared to almost negligible one in bulk materials.  For example,
10-layer Ta(110) exhibits a large anisotropy of the EYP of 82\%
compared to 0.2\% in bulk Ta.  The anisotropy of 44\% in Pt(001) is
much larger than 0.4\% in Pt bulk.  It has to be noted that for all
other films the maximal value of the EYP is also obtained if the SQA
is pointing perpendicular to the films.

However, in the hcp case the EYP shows a small anisotropy in thin
films compared to the large value in bulk materials.  For instance,
the EYP of bulk Os shows an anisotropy of 59\%,\cite{zimmermann12} while it
shows only a value of 7\% in 10-layer Os(0001) film.  The 10-layer
Hf(0001) film shows only 14\% of anisotropy of the EYP even though
there are surface states at the Fermi surface.  This is very small
compared to the gigantic value of 830\% of anisotropy of $b^2$ in bulk
Hf.\cite{zimmermann12} In Ref.~\onlinecite{zimmermann12,mokrousov13},
we demonstrated that the spin-flip part of SOC depends strongly on the
spin-quantization axis, possibly vanishing for one direction of the
SQA while being maximal for another. In a 
rough approximation, we can imagine that the Fermi surface of thin
films is constructed by the intersection of the Fermi surface of the
bulk sample with a number of planes that are parallel to the film
surface, with an inter-plane distance determined by the finite-size
quantization of crystal momentum in the direction perpendicular to the
film.  Therefore, the spin-flip hot spots or hot areas
that are formed at certain points in the bulk Fermi surface, e.g. at the hexagonal Brillouin
zone edge,\cite{zimmermann12} do not show in hcp(0001) film geometry
unless the film becomes thick enough and the intersecting planes dense
enough to capture these parts of the bulk Brillouin zone.

\begin{widetext}
\begin{table*}
\begin{tabular}{|c|c|c|c|c|c|c|c|c|c|c|}
\hline
Metal & $b^2$ & \multirow{2}{*}{$\mathcal{A}[b^2]$} & $T_{\rm p}$ (ps at.\%) & $T_{\rm sf}$ (ps at.\%) & $T_{\rm sf}^{-1}/T_{\rm p}^{-1}$ & \multirow{2}{*}{$\mathcal{A}[T_{\rm sf}^{-1}]$} & $\kappa_{yx}^s$ & $\kappa_{xx}$ & $\alpha=\kappa_{yx}^{s}/\kappa_{xx}$ & surf. \\ 
(surface) & $\hat{s}\parallel z$ & & $\hat{s} \parallel z$ & $\hat{s}\parallel z$ & $\hat{s}\parallel z$ &  & $({\rm m}\Omega cm)^{-1}$ & $({\rm \mu}\Omega cm)^{-1}$ & (\%) & states \\ \hline
Hf(0001) hcp& 0.143 & 14\% & 0.120 & 0.372 & 0.322 & 97\% & 29.00 & 12.67 & 0.228 & y \\ \hline
Ta(110) bcc& 0.117 & 82\% & 0.146 & 0.676 & 0.215 & 57\% & $-12.90$ & 22.98 & $-$0.056 & n \\ \hline
Ta(001) bcc& 0.103 & 24\% & 0.319 & 1.836 & 0.174 & 13\% & 32.91 & 17.44 & 0.188 & n \\ \hline
W(110) bcc& 0.229 & 57\% & 0.085 & 0.220 & 0.384 & 71\% & 13.41 & 11.99 & 0.111 & y \\ \hline
W(001) bcc& 0.294 & 37\% & 0.088 & 0.208 & 0.422 & 27\% & 0.77 & 1.12 & 0.069 & y \\ \hline
Re(0001) hcp& 0.187 & 7\% & 0.108 & 0.280 & 0.386 & 8\% & 7.85 & 10.13 & 0.077 & n \\ \hline
Os(0001) hcp& 0.183 & 7\% & 0.097 & 0.248 & 0.394 & 10\% & $-$12.59 & 18.09 & $-$0.069 & n \\ \hline
Ir(111) fcc& 0.182 & 21\% & 0.168 & 0.504 & 0.332 & 3\% & $-$26.91 & 27.22 & $-$0.098 & n \\ \hline
Ir(001) fcc& 0.268 & 46\% & 0.123 & 0.300 & 0.409 & 20\% & $-$5.42 & 10.34 & $-$0.052 & n \\ \hline
Pt(111) fcc& 0.168 & 37\% & 0.392 & 1.078 & 0.363 & 12\% & 45.89 & 30.42 & 0.150 & n \\ \hline
Pt(001) fcc& 0.286 & 44\% & 0.411 & 0.986 & 0.416 & 7\% & $-$17.14 & 16.41 & $-$0.104 & n \\ \hline
Au(111) fcc& 0.036 & 11\% & 0.166 & 2.282 & 0.072 & 3\% & $-$0.80 & 474.43 & $-$0.002 & y\\ \hline
Au(001) fcc& 0.065 & 50\% & 0.160 & 5.974 & 0.026 & 48\% & $-$71.03 & 217.28 & $-$0.326 & n \\ \hline
Cu(111) fcc& 0.0016 & 11\% & 0.175 & 43.40 & 0.004 & 94\% & $-$107.18 & 529.64 & $-$0.020 & y \\ \hline
Cu(001) fcc& 0.0024 & 29\% & 0.159 & 515.0 & 0.0003 & 24\% & $-$217.85 & 574.34 & $-$0.037 & n \\ \hline
\end{tabular}
\caption{Spin relaxation and spin Hall conductivity for 5{\it d}
  transition-metal 10-layer films, as well as Au and Cu, in different
  orientations. Ta and W are in the bcc structure, Ir, Pr, Au and Cu
  in the fcc structure, and Hf, Re and Os in the hcp structure
  (indicated in the first column). From left to right: the
  Elliott-Yafet parameter $b^2$ with the spin-quantization axis along
  the film normal $z$-axis, and its anisotropy; the momentum-relaxation time $T_{\rm p}$ and the
  spin-flip relaxation time $T_{\rm sf}$ as well as the ratio between the
  spin-flip and the momentum-relaxation rate with the
  spin-quantization axis $\hat s\parallel z$ and the anisotropy of
  the spin-flip relaxation rate; the transverse spin conductivity in
  $({\rm m}\Omega cm)^{-1}$, the charge conductivity in $(\mu\Omega
  cm)^{-1}$, the spin Hall angle $\alpha$, and the existence of
  surface states at \EF~(yes 'y' or no 'n').}
\label{alldata}
\end{table*}
\end{widetext}

\subsection{Spin relaxation due to self-adatom impurity \label{secspinrelax}}

In this section we discuss our results on the spin-relaxation process with self-adatoms as a source of scattering. 
The reason to choose adatom defects is that these naturally occur at any metal surface and additionally they comprise a reasonable generic model for surface
roughness. Other scattering mechanisms (different defects or phonons
at high temperature) would, of course, cause additional spin
relaxation.

First of all, it is interesting to compare the quantities
$b^2_{\vc{k}}$ and $\vc{k}$-dependent spin-flip relaxation rate
$T_{\rm sf}(\vc{k})^{-1}$ distributed on the Fermi surfaces, because in the
spirit of the Elliott approximation one expects a correlation between
them.  In the bottom of Fig.~\ref{bsqandT_WOs}, the distribution of
spin-flip relaxation rate on the Fermi surface of 10-layer W(110) (left)
and Os(0001) (right) films with the SQA along $z$-direction are
shown in comparison to $b^2_{\vc{k}}$ shown in the same figure.  The
{\bf k}-dependence of the spin-flip relaxation rate is obtained from the
scattering rate in Eq.~(\ref{krelaxtime}) as
$T_{\rm sf}(\vc{k})^{-1}=(1/2)\left(\left(\tau_{\vc{k}}^{+-}\right)^{-1}+\left(\tau_{\vc{k}}^{-+}\right)^{-1}\right)$.
Inspecting the color code of the two figures does not reveal a direct
$\vc{k}$-dependent one-to-one correspondence between $b^2_{\vc{k}}$
and $T_{\rm sf}(\vc{k})^{-1}$.  For instance, in W(110) the spin-relaxation
rate at $\vc{k}$-points belonging to surface states is very high,
while the value of $b^2_{\vc{k}}$ is not always high at the same
positions.  This also leads to a difference in the anisotropy value of
$b^2$ and $T_{\rm sf}^{-1}$ that will be discussed later.  A lack 
of direct correspondence between the values of $b^2_{\vc{k}}$ and $T_{\rm sf}(\vc{k})^{-1}$ is also
found in Os(0001), where we see that the value of $b^2_{\vc{k}}$ at
band crossings becomes almost maximal, while the value of
$T_{\rm sf}(\vc{k})^{-1}$ at the same points is moderate compared to the
half-rings on the outer part of the Brillouin zone.  Our conclusion is
that the Elliott approximation is too crude to give a correct
impression about the $\vc{k}$-dependent spin relaxation, but, as we
see below, it is qualitatively good for the $\vc{k}$-averaged
quantities that are anyhow the ones measured by experiment.

Averaging over the Fermi surface we obtain the spin-flip relaxation time
$T_{\rm sf}$ as well as the momentum-relaxation time $T_{\rm p}$.  The
calculated results in ps-at.\% for 10-layer films are presented in
columns 3 and 4 of Table~\ref{alldata}.  One can see that 
the systems with surface
states, i.e. 10-layer W(001) and W(110) films, have short spin-flip and
momentum-relaxation time as compared to others.  
For example, 10-layer W(110) film shows $T_{\rm p}$ of 0.085 ps-at\% and $T_{\rm sf}$ of 0.220 ps-at\% which are shorter than $T_{\rm p}$ of 0.146 ps-at\%
and $T_{\rm sf}$ of 0.676 ps-at\% in 10-layer Ta(110) film.  This is
intuitively expected since the scattering takes place at the adatoms,
with which surface states overlap more strongly than bulk states. The
momentum-relaxation time of other 5{\it d} thin films is of the
order of 0.1 to 0.4 ps-at.\% and the spin-flip relaxation time is of the order of 0.3 to 1.8 ps-at.\%.  
Comparing to the Au and Cu thin fims,
the momentum-relaxation time of 5{\it d} transition-metal thin films
is in the same order of magnitude, however, the spin-flip relaxation time
in 5{\it d} films is smaller by two or three orders of magnitude.  It
should be noted that the short spin-flip relaxation time of 5{\it d} thin
films is also influenced strongly by the SOC of self-adatom impurity
itself.

An interesting question arises: how good is the Elliott approximation, $T_{\rm sf}^{-1}/T_{\rm p}^{-1}\propto b^2$. 
To answer this, we report the ratio $T_{\rm sf}^{-1}/T_{\rm p}^{-1} $ for 10-layer films and summarized in Table~\ref{alldata}.  
In addition, a comparison between the ratio $T_{\rm sf}^{-1}/T_{\rm p}^{-1}$  and the EYP for a larger number of systems is shown in
Fig.~\ref{comparebandspinrelax}.  Interestingly, although the value
of spin-flip and momentum-relaxation rate varies very much among 5{\it d}
transition-metal thin films, the ratio between them scales
linearly, to a reasonable approximation, with the spin-mixing
parameter.  This qualitative result is not clear {\it a priori}, since the Elliott
approximation is based on the assumption of small values of
$b^2_{\vc{k}}$ and also neglects the form of the scattering potential that
should be too crude an approximation for transition-metal adatoms.

\begin{figure}
\includegraphics[scale=0.34,trim= 40 1 20 20,clip=true,angle=270]{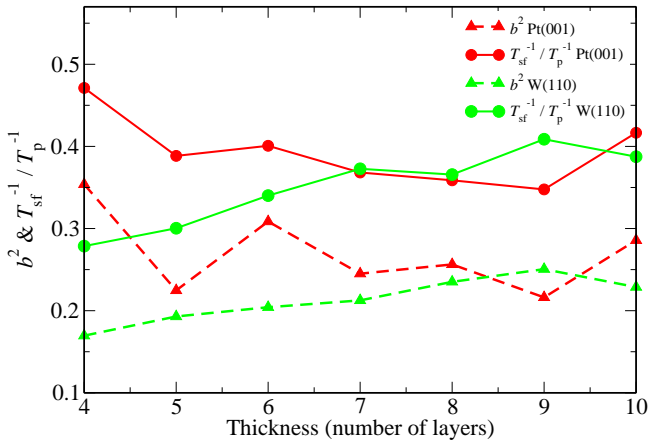} 
\caption{\small (Color online) Thickness dependence of the ratio $T_{\rm sf}^{-1}/T_{\rm
  p}^{-1}$ in Pt(001) and W(110) thin films with the self-adatom impurities.}
\label{T1layerPtW}
\end{figure}

We can also examine the correlation between the ratio $T_{\rm sf}^{-1}/T_{\rm p}^{-1}$ and the $b^2$ as a
function of the film-thickness.  In fact, in our recent work on
W(001),\cite{long13} we showed that the overall oscillating trend of
$T_{\rm sf}^{-1}/T_{\rm p}^{-1}$ with the film thickness corresponds well to
an oscillation of the EYP.  In Fig.~\ref{T1layerPtW} we show a similar
plot where the ratio $T_{\rm sf}^{-1}/T_{\rm p}^{-1}$ (solid lines) for Pt(001) (red)
and W(110) (green) as well as the EYP (dashed lines) are shown as a
function of the film thickness up to 10 layers (the SQA is taken
perpendicular to the film).  These two examples are chosen owing to
the fact that in Pt(001) both $T_{\rm sf}^{-1}/T_{\rm p}^{-1}$ and the EYP
show a oscillatory behavior.  In constrast, in W(110) both quantities
show an increasing behavior with increasing the film-thickness.  Of
course there is no one-to-one correspondence between $T_{\rm sf}^{-1}/T_{\rm p}^{-1}$ and the EYP in both films, but qualitatively they
show the same trends.  Once again, we can see the qualitative validity
of the Elliott approximation.

\begin{figure}
\includegraphics[scale=0.37,trim= 1 1 1 1,clip=true]{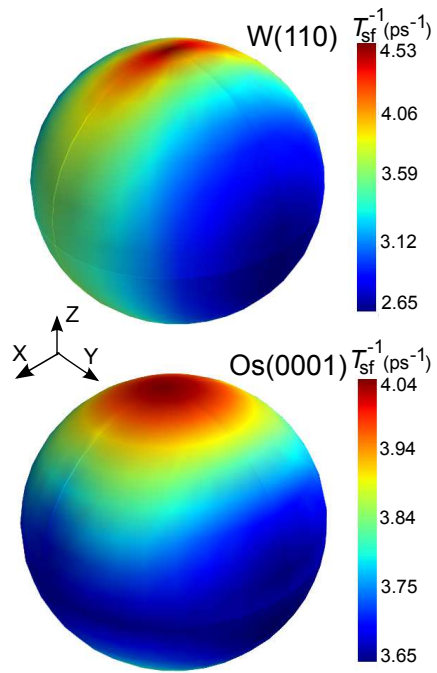} \\
\caption{\small (Color online) The spin-flip relaxation rate $T_{\rm sf}^{-1}$ in ps$^{-1}$/at.\%  of 10-layer W(110) film with W adatom and of
  10-layer Os(0001) film with Os adatom as scatterers as a function of
  the spin-quantization axis $\hat s$ on the unit
  sphere.}
\label{anitauWHf}
\end{figure}  

\begin{figure}
\includegraphics[scale=0.37,trim= 1 1 1 1,clip=true]{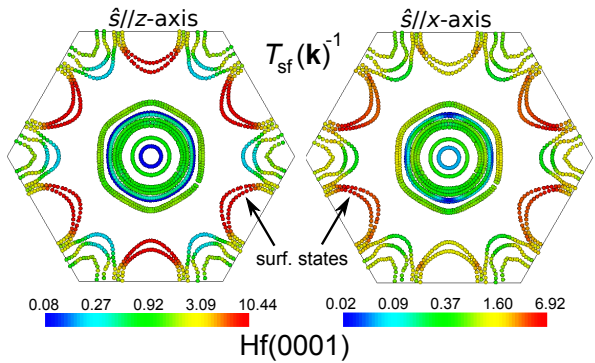} \\
\caption{\small (Color online) The distribution of spin-flip relaxation
  rate $T_{\rm sf}({\vc k})^{-1}$in ps$^{-1}$/at.\% on the Fermi surface of 10-layer Hf(0001)
  film with Hf adatom defect in the cases of the spin-quantization
  axis $\hat{s}\parallel z$ (left) and $\hat{s}\parallel x$
  (right). Surface states are indicated by arrows.}
\label{taukHf}
\end{figure}  

The anisotropy of the spin-flip relaxation rate is also investigated in
relation to the crystal symmetry by changing the spin-polarization
direction.  In Fig.~\ref{anitauWHf}, we show $T_{\rm sf}^{-1}(\hat{s})$ for
10-layer W(110) and Os(0001) films with the spin-quantization axis
$\hat s$ on the unit sphere.  Similarly to the case of the EYP, the
maximum value of $T_{\rm sf}^{-1}$ is obtained for $\hat s$ out of plane.
However, there is no one-to-one correspondence between the anisotropy of spin-flip relaxation rate and the EYP.

The calculated $\mathcal{A}\left[T_{\rm sf}^{-1}\right]$ in 10-layer films is
also shown in Table ~\ref{alldata}.  The anisotropy of 10-layer
Cu(111), Cu(001), Au(111) and Au(001) films is also calculated for
comparison.  It can be seen that in the films with surface states,
such as 10-layer W(110), Hf(0001), Au(111) and Cu(111) films with
self-adatom impurity, the anisotropy value of the spin-flip relaxation rate is
surprisingly higher as compared to that of the EYP.  In particular, in
10-layer Hf(0001) film, it reaches as mush as 97\%.  We can analyze
this by examining $T_{\rm sf}({\vc{k}})^{-1}$ on the Fermi surface of
10-layer Hf(0001) film with self-adatoms as scatterers in
Fig.~\ref{taukHf}.  Indeed, the $T_{\rm sf}^{-1}(\vc{k})$ at the surface
states with $\hat{s}\parallel z$ exhibits high values and it is much
lower for $\hat{s}\parallel x$.  As shown in
Fig.~\ref{bsqandT_WOs} for W(110) with W adatom impurity,
$T_{\rm sf}(\vc{k})^{-1}$ has also very high values at the surface states.
We can infer that the anisotropy of the spin-flip relaxation rate is highly
increased due to the Rashba surface states because of their
preferential, $\vc{k}$-dependent spin polarization (i.e. their Rashba
character) and because, as surface states, they have a strong overlap
with the adatom scatterers.

The anisotropy value of the spin-flip relaxation rate in other 10-layer thin
films with self-adatom impurities as scatterers and without surface
states is comparable to that of the EYP.  The cubic systems show a
high anisotropy of spin-flip relaxation rate such as 57\% in Ta(110) with
Ta adatom impurity or 48\% in Au(001) with Au adatom impurity.
However, similar to the anisotropy of EYP, the hcp thin films show a
small value $\mathcal{A}\left[T_{\rm sf}^{-1}\right]= 10\%$ in Os(0001) and
8\% in Re(0001).  This result is expected and it can be explained in a
similar way for the anisotropy of the EYP.  As discussed in
Ref.~\onlinecite{zimmermann12,long13,long13b,mokrousov13}, there is no theoretical limit 
on the value of anisotropy, and as a consequence, this value depends very much on the material parameters.

\subsection{Spin Hall conductivity \label{secspinhall}}

We proceed with an investigation of the extrinsic spin Hall
conductivity due to skew-scattering off self-adatoms. 
Assuming that we set $x$-direction as the charge current direction and $z$-direction as the spin-polarization direction, the central experimentally accessible quantity in the spin Hall effect is the spin
Hall angle $\alpha=\kappa_{yx}^s/\kappa_{xx}$ relating the transverse
spin current to the longitudinal charge current. 
$\kappa_{yx}^s$ and $\kappa_{xx}$ are the off-diagonal elements of the spin-conductivity tensor (Eq.~\ref{eq:scond}) and the diagonal elements of
the charge-conductivity tensor (Eq.~\ref{eq:ccond}), respectively.

\begin{figure}
\includegraphics[scale=0.34,trim= 40 1 20 20,clip=true,angle=270]{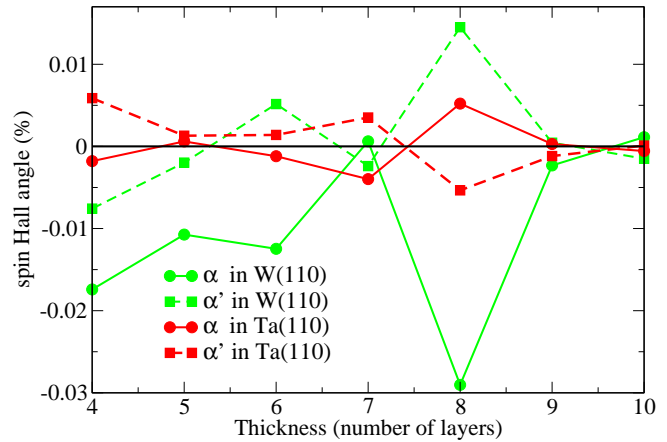} 
\caption{ (Color online) Spin Hall angles $\alpha=\kappa_{yx}^s/\kappa_{xx}$ and $\alpha'=-\kappa_{xy}^s/\kappa_{yy}$ of W(110) with W
  adatom impurity and Ta(110) with Ta adatom impurity as a function of
  the film thickness.}
\label{spinhallWTa}
\end{figure}

First we note on the anisotropy of spin Hall angle with respect to the current direction in bcc (110) films.
From the conductivity tensor, in principle, we can have two definitions of spin Hall angle depending on the direction of the longitudinal current.  
The spin Hall angle defined above corresponds to the electric field applied along the $x$-axis.  
We can also apply an electric field along the $y$-axis, which will result in the following value for the
spin Hall angle $\alpha'=-\kappa^s_{xy}/\kappa_{yy}$ as measured experimentally.
In case of a high in-plane symmetry, which is the case for~e.g.~W(001) films, $\alpha=\alpha'$, while when the in-plane symmetry is lowered,~e.g.~W(110) and Ta(110) films, the spin Hall angles $\alpha$ and $\alpha'$ can be different from each other.  In Fig.~\ref{spinhallWTa},
the spin Hall angles $\alpha$ and $\alpha'$ of W(110) with W adatom as scatterer and Ta(110) with Ta adatom as scatterer are plotted as a function of the film-thickness.  
It can be seen that the spin Hall angles can vary very much.  
In many cases, even the sign of the SHA $\alpha$ and $\alpha'$ is
different.  This leads to a
high anisotropy effect of the spin Hall angle in such systems with
respect to the current direction.  As seen in
Fig.~\ref{spinhallWTa}, the anisotropy can reach up to 300\% in
8-layer W(110) film or 200\% in 7-layer Ta(110) film.
The high values of anisotropy of spin Hall angle come from the anisotropy in both spin and charge current.   
One should stress that the anisotropy of SHA here is with respect to the current direction in the film and it is in contrast to the EYP and spin relaxation case where we observed an anisotropy with respect to the spin-polarization direction $\hat s$.

\begin{figure}
\includegraphics[scale=0.34,trim= 40 1 20 20,clip=true,angle=270]{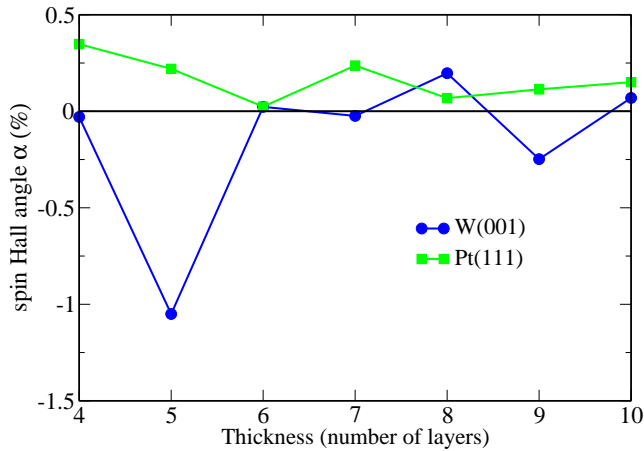} 
\caption{\small (Color online) Spin Hall angle $\alpha$ of W(001) with W adatom defect and Pt(111) with Pt adatom defect as a function of the film thickness.}
\label{spinhallPtW}
\end{figure}

It is also rewarding to observe that the spin Hall angle oscillates
with increasing film thickness. 
In Fig.~\ref{spinhallPtW}, the spin Hall angle $\alpha=\kappa_{yx}^s/\kappa_{xx}$ is shown as a function of film thickness for W(001) and
Pt(111) with self-adatom impurities.  From the definition of the spin Hall angle, one can expect an independence of $\alpha$ on the film thickness. 
However, in both cases, the spin Hall angle shows an oscillatory dependence on the film thickness, indicating
quantum confinement effect. \cite{herschbach12} We have previously
seen oscillatory effects also in the EYP and the spin-flip relaxation rate
of W(001) films and Pt(111) films as a function of thickness shown in
Ref.~\onlinecite{long13} as well as in the present work.  However, the variation
curve of spin Hall angle is different from that of the EYP or the spin
relaxation.  In fact a correspondence between the spin-flip relaxation rate
and the spin Hall angle is not expected: as can be seen from the
scattering-in term in Eq.~(\ref{lambda}), the spin Hall conductivity
is determined by the contributions of the spin-conserving and
spin-flip probability, while the spin-flip relaxation is determined only by
the spin-flip probability.

\begin{figure}
\includegraphics[scale=0.37]{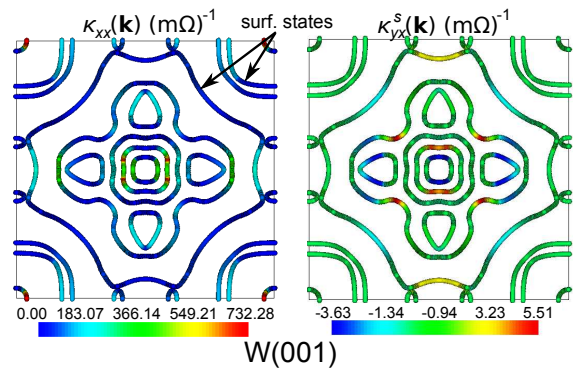} 
\caption{(Color online) The distribution of the charge conductivity
  $\kappa_{xx}(\vc{k})$ and the transverse spin conductivity
  $\kappa_{yx}^s(\vc{k})$ in $(\rm milliohm)^{-1}$ on the Fermi surface
  of 10-layer W(001) film with 1\% self-adatom impurities.}
\label{sigmaW001}
\end{figure}

We investigate the effect in more detail by plotting the distribution
of the charge and transverse spin conductivty on the Fermi surface of
10-layer W(001) film with 1\% self-adatom impurities in
Fig.~\ref{sigmaW001}.  Quantities $\kappa_{xx}(\vc{k})$ and $\kappa_{yx}^{s}(\vc{k})$ are simply defined as $\vc{k}$-distributions of the charge- and
spin-conductivities on the Fermi surface,
$\kappa_{xx(yx)}^{(s)}=\int_{\rm
  FS}d\vc{k}\cdot\kappa_{xx(yx)}^{(s)}\left(\vc{k}\right)$
  and the units of $\kappa(\vc{k})$ is
$(\rm milliohm)^{-1}$ (c.f. also Eqs.~\ref{eq:ccond} and \ref{eq:scond}).  Obviously, the bulk-like states carry most
of the longitudinal charge current and the surface states do not,
i.e. the charge conductivity is very low at surface states.  On the
contrary, one can expect that the surface states can carry a
transverse spin current due to the very strong
scattering.\cite{herschbach12} However, it is not the case for this
system.  Large transverse spin conductivities are also seen in the bulk-like
states.

Table~\ref{alldata} summarizes the charge and spin conductivity as
well as the spin Hall angle $\alpha$ of 5{\it d} 10-layer films together with
10-layer Au and Cu.  The impurity concentration is 1\% for all
calculations.  The values for W(110) and Ta(110) films are illustrated with
$\alpha=\kappa_{yx}^s/\kappa_{xx}$.  A first impression is that
the spin Hall angles are rather small in magnitude for all systems.
Moreover, they are quite different in magnitude and sign when
changing the material.  Thin films with Rashba surface states at the Fermi
surface are expected to have a large spin Hall angle.  However, our
calculations show rather small spin Hall angles in such films.  For
example, 10-layer W(001) film with W adatom impurity shows only a value
of 0.06\% for spin Hall angle, while its value in 10-layer Au(111) film with Au
adatom impurity is very small constituting only $-$0.0017\%.  Comparing between Cu
films and 5$d$ transition-metal films with different strength of SOC,
we can observe no clear trend in the magnitude of the
spin Hall angle with increasing the SOC strength.

\section{Summary}

In this work, we studied the consequences of spin-dependent scattering
in non-magnetic metallic thin films.  We particularly focused on
the effects of the spin relaxation induced by the Elliott-Yafet
mechanism as well as the extrinsic spin Hall transport due to the
skew-scattering for 5{\it d} transition-metal thin films with
self-adatom impurity in comparison with Au and Cu thin films.

The Elliott-Yafet parameter and the spin relaxation are systematically
examined as functions of the film thickness up to 10 layers as well as
the crystallographic orientation of the film.  The overall trends are in
qualitative agreement with the Elliott approximation.  Quantitatively,
due to strong spin-flip scattering and complicated electronic
structure in {\it d}-orbital materials, the spin-flip relaxation time of 5{\it
  d} transition-metals with self-adatom impurity is roughly about few hundred nanoseconds at atomic percent which is two or three order of magnitude shorter
than that of Cu and Au thin films.  

Owing to the reduced dimensionality, the anisotropy of the spin-mixing
parameter and the spin-flip relaxation rate in thin films is different
from that in bulk metals, but not in a universal manner.  For
cubic crystal structures, the anisotropy significantly increases in
thin films compared to that in bulk systems, because of the
crystal-symmetry reduction.  On the contrary, in hcp materials where
the symmetry in bulk is anyhow low, the anisotropy value in bulk is
quite large and in all studied cases higher than the value in thin
films, as a result of the Fermi surface formation.  Furthermore, we
find that the presence of Rashba surface states plays a crucial role in
the spin relaxation.  E.g., the anisotropy of spin-flip relaxation rate
reaches a value of 97\% in 10-layer Hf(0001) or 71\% in 10-layer
W(110) film.

The longitudinal charge conductivity and the transverse spin Hall
current for 10-layer thin films with 1\% self-adatom impurities are
calculated by means of the self-consistent Boltzmann equation. 
The spin Hall angle found to strongly vary in $5d$ films with respect to the material but also with respect to film thickness and orientation.

\section{acknowledgments} 
This work was financially supported by Deutsche Forschungsgemeinschaft
projects MO 1731/3-1 and SPP 1538 SpinCaT, and the HGF-YIG NG-513
project of the Helmholtz Gemeinschaft.  The authors are grateful to
Gustav Bihlmayer, Rudolf Zeller, and Peter Dederichs, for fruitful
discussions. We acknowledge computing time on the supercomputers JUQEEN and JUROPA at J\"ulich Supercomputing Center and JARA-HPC of RWTH Aachen University.


\bibliography{apssamp}

\end{document}